\documentclass[superscriptaddress,twocolumn,showpacs,prl,amsmath,amssymb]{revtex4}
\usepackage{graphicx}% Include figure files
\usepackage{bm}% bold math

\usepackage[varg]{txfonts}

\def\XXint#1#2#3{{\setbox0=\hbox{$#1{#2#3}{\int}$}
     \vcenter{\hbox{$#2#3$}}\kern-.5\wd0}}

\begin{document}

\title{The internal Josephson effect in a Fermi gas near a Feshbach resonance} 

\author{V. M. Galitski} \affiliation{Condensed Matter Theory Center, Department
    of Physics, University of Maryland, College Park, MD 20742-4111}
\begin{abstract}
We consider a two-component system of Fermi atoms and molecular bosons
in the vicinity of a Feshbash resonance.  We derive an effective
action for the system, which has a term describing coherent tunneling
of the molecular bosons into Cooper pairs and {\em vice versa}.  In
the equilibrium state, global phase coherence may be destroyed by
thermal or quantum phase fluctuations. In the non-equilibrium
regime, the system may show an internal AC Josephson effect leading
to real time oscillations in the number of molecular bosons.
\end{abstract}

\maketitle
\vspace*{-0.17in}

{\em Introduction.}  Trapped dilute cold Fermi systems \cite{PT,Cho}
are one of the most exciting areas of research in modern condensed
matter physics.  This field offers a great variety of novel phenomena
\cite{oscexp,JinFC,KetterleFC,JinNature,JinLifetime,Jina(B)} and
presents serious challenges for both experimentalists and theorists.
The magnetic-field induced Feshbach resonance \cite{Feshbach} provides
an unprecedented degree of control over the inter-particle
interactions as well as the rate at which the interactions are
changed.  By sweeping magnetic field, one can tune the sign and the
strength of the interaction \cite{Jina(B)} and experimentally access
the BCS-BEC crossover physics \cite{Noz}. Apart from this, by
adjusting the sweeping rate, one can drive the system into various
non-equilibrium states in which different bosonic and fermionic
species co-exist and interact with each other.

Interaction between fermions consists of two channels: an inelastic
channel involving the formation of a molecular state of two fermions
and a resonant elastic scattering channel.  Typically, the elastic
scattering length is an unremarkable function of the applied field.
However, in the vicinity of a certain value of the field $B_0$, the
scattering length changes dramatically \cite{Feshbach,Jina(B)}.  On
the BEC side of the Feshbach resonance, the molecular energy level is
located below the continuum states and the molecular state is stable.
On the BCS side of the resonance, high magnetic field breaks up the
molecular states and one would expect that the molecular component
vanish.  However, recent experimental results \cite{JinLifetime} and
theoretical works \cite{Griffin,Falco} suggest that even on the BCS
side of the resonance, the fraction of molecular states is non-zero.
One may assume the existence of an energy barrier between the
molecular state and the continuum and so the molecular state on the
BCS side of the resonance may actually be metastable.  This may happen
in part due to a stabilizing effect of the Fermi liquid coupled to the
molecular component. In turn, the molecular component affects the
fermionic degrees of freedom renormalizing effective fermion-fermion
interactions \cite{Griffin,Stajic}.

At low temperatures, attractive interaction between fermions should
lead to the formation of Cooper pairs \cite{Bohn,JinFC,KetterleFC}.
The condensate of Cooper pairs would co-exist and interact with the
system of preformed molecular bosons if the life-time of the molecular
state is long enough.  It is interesting to describe the co-existence
of the two coherent states, and this issue is the main subject of the
present work.  In this Letter, we study theoretically a Fermi-Bose
mixture and discuss quasiequilibrium and non-equilibrium properties of
the system.

We start with an effective low energy Hamiltonian describing the
system near the resonance (on the high-field BCS side)
\cite{Holland,Stoof}. In this Hamiltonian, the resonant and
non-resonant processes are separated by introducing an effective
Feshbach coupling of the fermion component to the metastable molecular
field.  We integrate out the one-particle fermionic degrees of freedom
and explicitly derive an effective action for the system. This action
contains a term which can be interpreted as an internal Josepshon
tunneling \cite{Leggett} of molecular bosons into Cooper pairs and
{\em vice versa.}  The action also contains a ``charging energy''
term, describing phase diffusion. In the quasiequilibrium regime,
phase coherence between the two states can be destroyed by either
temperature or quantum phase fluctuations. We explicitly derive the
crossover line, which is a function of the number of particles in the
system $N$. In the limit $N \to \infty$, the internal DC Josephson
effect is dominant and global phase coherence is established in the
system.

We also discuss a non-equilibrium situation by considering a
non-adiabatic sweep of the magnetic field across the resonance.  We
predict that under certain circumstances, the system may exhibit
Josephson AC oscillations. In this case, the numbers of molecular
bosons and Cooper pairs oscillate in real time, slowly relaxing to a
stationary state, which depends on the magnetic field sweeping rate.
In addition, we discuss relevant time-scales at which such a
relaxation takes place.

{\em Effective action.} Let us consider a system of Fermi atoms with
two hyperfine states labeled with the index $\sigma = \uparrow,\,
\downarrow$.  The effective Hamiltonian for the system in the vicinity
of the Feshbach resonance has the following form \cite{Holland}:
\begin{eqnarray}
\label{H}
\nonumber
{\cal H} - \mu N = \int d^3 {\bf r} &&\!\!\!\!\! \Biggl\{ \psi^{\dagger}_\sigma\left( {\bf
  r} \right) 
\left[ - {\nabla^2 \over 2 m} - \mu \right]
\psi_\sigma\left( {\bf
  r} \right) \\
\nonumber
&& - { U \over 2}
\psi^{\dagger}_\sigma\left( {\bf r} \right) \psi^{\dagger}_{-\sigma}\left( {\bf r} \right) 
\psi_{-\sigma}\left( {\bf  r} \right)
\psi_\sigma\left( {\bf r} \right)  \\
&& + \phi^\dagger \left( {\bf r} \right) 
\left[ - {\nabla^2 \over 2 M} + \delta -  2\mu \right]
\phi \left( {\bf r} \right) \\
\nonumber
&&+ g \left[ \phi^\dagger \left( {\bf r} \right) \psi_{\sigma}\left( {\bf
  r} \right) \psi_{-\sigma}\left( {\bf
  r} \right)  + \mbox{h.c.} \right] \Biggr\},
\end{eqnarray}
where $\psi$ and $\phi$ are field operators for fermions and molecular
bosons respectively, $U$ is the fermion-fermion interaction strength,
$\delta \sim 2 \mu_{\rm B} \left( B - B_0 \right)$ is the detuning
from the resonance, $g$ is the Feshbach coupling between molecular
bosons and fermions (which is connected with the width of the
resonance $\Delta B$ \cite{Stoof}), and $m$ and $M$ are fermionic and
molecular masses respectively.

The grand partition function for the system is
\begin{equation}
\label{Z}
Z_G = {\rm Tr}\, \exp\left[- \beta \left( {\cal H} - \mu N \right)
\right],
\end{equation} 
where the trace is taken over the fermionic and bosonic degrees of
freedom. Now let us introduce the Hubbard-Stratonovich field $\Delta$
to decouple the four-fermion term $H_U$ in the action, which can be
re-written in the following form
\begin{eqnarray}
\label{H4}
\nonumber
&& T_{\tau}\, \exp \left[ - \int\limits_0^{\beta} d\tau H_U (\tau)
\right] = \, \int {\cal D}^2\Delta(x)  
 e^{\left[ - U^{-1} \int dx  \left| \Delta (x) \right|^2 \right]} 
\\
&&\,\,\,\,\,\,\,\,\,\,\,\,\,\,\,\,\,\,
\times T_{\tau} \exp\left[ - \int dx \left\{
\Delta(x) \psi^{\dagger}_\sigma (x) \psi^{^\dagger}_{-\sigma}(x) +
\mbox{h.c.} \right\} \right],
\end{eqnarray}
where we introduced $x = \left( \tau, {\bf r} \right)$ for the sake of
brevity. Let us note that the last term in Eq.~(\ref{H4}) is very
similar to the Feshbach coupling term in Eq.~(\ref{H}).  To proceed
further, let us evaluate the Gaussian integral over the fermionic
degrees of freedom in the spirit of Ref.~\cite{AES}. We use the
following standard Nambu notations:
$$
\hat{\psi}(x) = {  \psi_\uparrow (x)  \choose \psi^\dagger_\downarrow
  (x)}.
$$
The Hubbard-Stratonovich and  molecular fields read
$$
\hat \Delta(x) = 
\left( \begin{array}{cc}
  0 & \Delta(x) \\
\Delta^*(x) & 0
\end{array} \right)\,\,\, \mbox{and} \,\,\,
\hat \phi(x) = 
\left( \begin{array}{cc}
  0 & \phi(x) \\
\phi^*(x) & 0
\end{array} \right).
$$
The Green's function $\hat G(x) = - \left\langle T_{\tau}\,
\hat\psi(x) \hat\psi^{\dagger}(x) \right\rangle$ is also a $2 \times
2$ matrix. In the Nambu notations, the grand partition function takes
the following form:
\begin{equation}
\label{Z2}
Z_G = \int {\cal D}^2 \Delta(x) \, {\cal D}^2 \phi(x) \exp \left\{ -
S_{\rm eff} \left[ \hat\Delta, \hat\phi \right] \right\},
\end{equation}
where the effective action can be written in the compact form:
\begin{eqnarray}
\label{S1}
\nonumber
S_{\rm eff}  \left[ \hat\Delta, \hat\phi \right] = && \!\!
\int\limits_0^\beta d\tau {\cal H}^{(0)}_{\phi}(\tau) + 
{1 \over U} \int dx \left| \Delta(x) \right|^2  \\
&& \!\!\!\! -
{\bf{\rm Tr}} \left[ \ln \hat G_0^{-1} + \ln \hat G^{-1} \right].
\end{eqnarray}
Where the trace is over the spatial and time variables and Nambu
indices and ${\cal H}^{(0)}_{\phi}$ is the Hamiltonian of the system
of molecular bosons (without the Feshbach coupling). The matrix
Green's function (which quite generally is a functional of the pairing
and bosonic fields) is the solution of the following equation:
\begin{eqnarray}
\label{G}
\nonumber
\left[ - {\partial \over \partial \tau} \hat{1}  - \left( - {\nabla^2 \over 2
    m} - \mu \right) \hat\sigma_3 - \hat\Delta(x) -
   g  \hat\phi(x)
    \right] \hat G \left(x,x' \right)\\ = \delta(t-t')\,\delta({\bf r} - {\bf
    r}') ,
\end{eqnarray}
where $\hat\sigma_{\alpha}$ is the Pauli matrix in the Nambu space.

In what follows, we consider a homogeneous order parameter. This
approximation is exact in the zero-dimensional case, {\em i.~e.}  when
the size of the system is smaller than the coherence length, otherwise
one should study Ginzburg-Landau equations taking into account the
spatial dependence of the fields.

Both the pairing field and the Bose-field are complex functions of the
imaginary time. Let us write them in the form:
$$
\Delta(\tau) = \Delta_0 e^{i \chi_1(\tau)} \,\,\, \mbox{and} \,\,\,
\phi(\tau) = \phi_0\, e^{i \chi_2(\tau)}
$$
and perform the following gauge transformation 
$$
\hat{\tilde\psi}(x) = \left(
\begin{array}{cc}
 e^{-i \chi_1(\tau) / 2} & 0\\
0 &  e^{i \chi_1(\tau) / 2}
\end{array} \right) \hat\psi(x).
$$
In this gauge, the Green's function reads (although the coordinate
dependences of the order parameters are neglected, the Green's
function depends on ${\bf r}$ and ${\bf r}'$):
\begin{eqnarray}
\label{G2}
\nonumber
\left[ - {\partial \over \partial \tau} \hat{1}  - \left( - {\nabla^2 \over 2
    m} - \mu \right) \hat\sigma_3 - \hat\Delta_0 -
   g \hat{\tilde{\phi}} (\tau) - {i \over 2} {\partial \chi_1 \over \partial
      \tau} \hat\sigma_3
    \right]\\
\times \hat{\tilde{G}} \left( \tau, \tau' ; {\bf r} - {\bf r}' \right)= \delta(t-t')\,\delta({\bf r} - {\bf
    r}'). 
\end{eqnarray}
Where the new bosonic field (the order parameter describing a
superfluid phase of molecular bosons) is given by
$$
 \hat{\tilde{\phi}} = \phi_0 
\left( \begin{array}{cc}
0 & e^{ i \left( \chi_2 - \chi_1 \right)} \\
 e^{ i \left( \chi_1 - \chi_2 \right)} & 0
\end{array} \right).
$$

Now, let us consider the case when the bosonic and dynamic terms in
the Green's function are small compared to the other terms. In this
limit, the effective action can be expanded with respect to the small
terms:
\begin{eqnarray}
\label{S2}
\nonumber
 S_{\rm eff}  \left[ \hat\Delta, \hat\phi \right] = 
&&\!\!\!\!\! -\,
{\bf{\rm Tr}} \left[ \ln \hat G_0^{-1} - \ln \hat G_{\rm GN}^{-1}
\right]  + 
{V \beta \over U} \Delta_0^2 \\
\nonumber
&&\!\!\!\!\! + \int\limits_0^\beta d\tau {\cal H}^{(0)}_{\phi}(\tau)  
 +\, {\bf{\rm Tr}} \left[  \hat G_{\rm GN} \left( g \hat\phi +   {i \over 2} {\partial \chi_1 \over \partial
      \tau} \hat\sigma_3 \right) \right] \nonumber \\
&&\!\!\!\!\! +\, {\bf{\rm Tr}} \left[  \hat G_{\rm GN} \left( g \hat\phi +   {i \over 2} {\partial \chi_1 \over \partial
      \tau} \hat\sigma_3 \right) \right]^2 + \ldots,
\end{eqnarray}
where $\hat G_{\rm GN}$ is the Gor'kov-Nambu Green's function matrix:
\begin{equation}
\label{hatG}
\hat{G}_{\rm GN} (\tau, {\bf r}) = 
\left(
\begin{array}{cc}
G(\tau, {\bf r}) & F(\tau, {\bf r}) \\
 F(\tau, {\bf r}) & - G(-\tau, -{\bf r})
\end{array}
\right).
\end{equation}
In the momentum representation, the components of the matrix have the
following well-known form \cite{AGD}:
\begin{equation}
\label{GF}
G\left( \varepsilon_n,{\bf p} \right)
= - {\xi_p + i \varepsilon_n \over \Delta_0^2 +
\xi_p^2 + \varepsilon^2 }\,\,\, \mbox{and} \,\,\,
 F \left( \varepsilon_n,{\bf p}
\right) = {\Delta_0 \over \Delta_0^2 +
\xi_p^2 + \varepsilon_n^2 },
\end{equation}
where $\epsilon_n = \left( 2 n + 1 \right) \pi T$ is the fermionic
Matsubara frequency.  The first two terms in action (\ref{S2}) pin the
mean-field value of the BCS order parameter, while the third term
controls the magnitude of the superfluid molecular field [in principle
one can include higher order terms with respect to $\phi$ in the
initial Hamiltonian (\ref{H}), describing interactions between
molecules; the tunneling terms in the action will remain the same in
this case].  The most interesting contributions come from the fourth
and fifth terms in Eq.~(\ref{S2}), which describe quantum phase
dynamics.  Using Eqs.~(\ref{S2}), (\ref{hatG}), and (\ref{GF}), we
obtain the first order correction to the mean-field action:
% $$
% {i \over 2} {\rm Tr\,} \left[ \hat G_{\rm GN} {\partial \chi_1 \over
%   \partial \tau} \hat{\sigma}_3 \right] =0
% $$
% and
\begin{equation}
\label{Jos}
\delta S_1 = {\rm Tr\,} \left[ \hat G_{\rm GN} {\partial \chi_1 \over \partial
  \tau} \hat{\sigma}_3 \right] =- J \int\limits_0^{\beta} d \tau 
\cos \left[ \chi_1\left( \tau \right) - \chi_2 \left( \tau \right)
\right],
\end{equation}
with 
\begin{equation}
\label{JJ}
J(T) = {3 \over 2}  N {g \phi_0(T) \Delta_0(T) \over \epsilon_{\rm F}}
\ln{\left[ 
\tilde{\epsilon} \over {\rm max}\, \left\{ \Delta_0(T), T \right\}
\right]},
\end{equation}
where $N$ is the number of particles in the Fermi system,
$\epsilon_{\rm F}$ is the Fermi energy, and $\tilde{\epsilon}$ is the
BCS high-energy cut-off (in our case $\tilde{\epsilon} \sim
\epsilon_{\rm F}$).  The leading contribution coming from the second
order correction term [the last term in Eq.~(\ref{S2})] generally has
a complicated non-local
% \begin{eqnarray}
% \label{dS2}
% \delta S_2 = && \!\!\!\! - {V \delta_{\rm F} \over 4}
% \int d \tau d \tau'  \dot{\theta}(\tau) \dot{\theta}(\tau')\\
% && \!\!\!\! \times \Biggl\{  \int d \xi_{\bf k} \Bigl[ 
%  G\left( {\bf k}, \tau - \tau' \right)
% G\left( {\bf k}, \tau' - \tau \right) \\
% &&  \!\!\!\! - F\left( {\bf k}, \tau - \tau' \right)
% F\left( {\bf k}, \tau' - \tau \right) \Bigr] \Biggr\}
% \end{eqnarray}
form. However, if the phase dynamics is slow enough at the time-scale
$\Delta_0^{-1}$, the action can be reduced to the familiar local form:
\begin{equation}
\label{Scharge}
\delta S_2 = {1 \over 8 E_{\rm C}} \int\limits_0^\beta 
d \tau \left( {\partial \chi_1 \over \partial \tau} \right)^2,
\end{equation}
where 
\begin{equation}
\label{Ec}
E_{\rm C}^{-1} (T) ={3 c(T) \over 2} {N \over  \epsilon_{\rm F}}
\end{equation}
and the function $c(T)$ is given by
\begin{equation}
\label{c}
c(T) = 
\int\limits_{- {\infty}}^\infty d\xi
\left\{ {\Delta_0^2(T) \over E^3} \tanh\left[ {E \over 2 T} 
\right] - {\xi^2 \over 4 T E^2} \cosh^{-2} \left[  {E \over 2 T}
\right] \right\},
\end{equation}
with $E = \sqrt{ \Delta_0^2(T) + \xi^2}$. This function has the
following asymptotic behavior at low and high temperatures: $c(T \ll
\Delta_0) = 1$ and $c(T \gg \Delta_0) \approx \pi \Delta_0(T)/T$.
Summarizing, let us present the main technical result of the present
work:
\begin{equation}
\label{SSS}
\delta S 
=  \int\limits_0^{\beta} d \tau \left[ - J
\cos \left[ \chi_1\left( \tau \right) - \chi_2 \left( \tau \right)
\right] + {1 \over 8 E_{\rm C} }   \left( {\partial \chi_1 \over
  \partial \tau} \right)^2 \right],
\end{equation}
where the parameters $J$ and $E_C$ are given by Eqs.~(\ref{JJ}) and
(\ref{Ec}).  The first term in action (\ref{SSS}) can be easily
recognized as a Josephson term, which describes coherent tunneling of
the molecular bosons into the condensate of fermionic Cooper pairs and
{\em vice versa}. Unlike in the conventional Josephson effect
\cite{Josephson}, the two quantum states are not separated by any
physical energy barrier being mixed up in the real space
\cite{Leggett}. The second term in Eq.~(\ref{SSS}) is the effective
``charging energy.'' This term describes quantum phase fluctuations.
In a neutral superfluid, phase dynamics is possible only due to the
finite size of the system (this is different from the case of a
superconducting junction, when quantum phase fluctuations are possible
in any system of finite capacitance).

{\em Quasiequilibrium regime.} Let us consider first a system of
Cooper pairs and preformed molecular bosons in the thermodynamic
equilibrium, {\em i.~e.}  $\dot\phi = 0$. In this state, the molecular
order parameter is directly related to the BCS order
parameter~\cite{eqphi}. The corresponding relation follows from the
condition $\left[ \phi , {\cal H} \right] = 0$ and Eq.~(\ref{H}):
\begin{equation}
\label{equil}
\phi_0 = {g \over 2 \mu - \delta } {\Delta_0 \over U}.
\end{equation}
\vspace*{-0.15in}
\begin{figure}[htbp]
\centering
\includegraphics[width=3.3in]{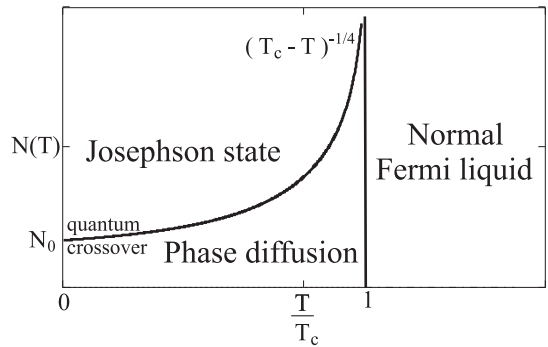}
\caption{ This diagram  describes a crossover between the phase
  coherent Josephson dynamics and the phase oscillating (Fock) regime.
  The curve shows the critical number of particles as a function of
  temperature at which the crossover occurs.  }
\label{fig1}
\end{figure}

In the quasiequilibrium regime, the internal Josephson effect tends to
establish phase coherence between the two condensates. However, even
at zero temperature, this coherence can be destroyed by quantum phase
fluctuations due to a finite mass term in the action (\ref{SSS}). A
crossover between the semiclassical Josephson dynamics and quantum
phase diffusion occurs when \cite{Raghavan} $J \sim E_{\rm C}$.  From
this condition and Eqs.~(\ref{JJ}), (\ref{Ec}), and (\ref{equil}) (at
zero temperature), we derive the following expression for the number
of particles in the Fermi liquid corresponding to the quantum
crossover point:
\begin{equation}
\label{qpt}
N_{\rm c} (0) \sim \left[ {\nu_{\rm F} U^2 \left( \delta - 2\mu \right) \over
  g^2} \right]^{1/2} e^{1 / \left( \nu_{\rm F} U \right)},
\end{equation}
where $\nu_{\rm F}$ is the density of states at the Fermi level and we
have used the BCS formula for the transition temperature with
$\epsilon_{\rm F}$ as a high-energy cut-off. We emphasize that the
effect can not be called a quantum phase transition, since the
transition is driven by finite size of the system and disappears in
the thermodynamic limit (the system is always in the phase coherent
state if $N = \infty$). At finite temperatures, the Josephson effect
is suppressed by thermal fluctuations.  The corresponding crossover
between the coherent state of two condensates and random phase
rotation can be estimated from the relation $J(T) \sim E_{\rm C}(T)$,
using Eqs.~(\ref{JJ}) and (\ref{Ec}):
\begin{equation}
\label{pt}
N_{\rm c} (T) \sim N_{\rm c} (0) \sqrt{c(T)} {\Delta_0 \over \Delta_0 (T)},
\end{equation}
where the function $c(T)$ is given by Eq.~(\ref{c}) and $\Delta_0(T)$
is the temperature dependent BCS gap \cite{AGD}. The $N_{\rm
  c}(T)$-dependence was evaluated numerically and the corresponding
curve is shown in Fig.~1.

{\em Non-equilibrium dynamics.} Let us consider a non-equilibrium
case, when a magnetic field is swept suddenly from the BEC to the BCS
side of the resonance. In this case one has a non-equilibrium
situation: the distribution function of quasiparticles $f({\bf k})$ is
not given by the familiar equilibrium distribution $f_0({\bf k}) =
\left[ 1 + \exp\left( \beta \sqrt{\xi_{\bf k}^2 + \Delta^2} \right)
\right]^{-1}$.  The technique used previously is formally valid only
if the time-scale at which the order parameter changes is much larger
than the quasiparticle relaxation rate $\tau_\Delta \gg
\tau_\epsilon$. This is definitely true in the close vicinity of the
transition point [$\Delta \ll T_c \sim \Delta(T=0)$]: $\tau_\Delta
\sim \tau_\varepsilon {T_c / \Delta} \gg \tau_\varepsilon$, where
$\tau_\epsilon^{-1}$ is the quasiparticle relaxation rate, which in a
clean Fermi liquid can be estimated as $\tau_\epsilon^{-1} \sim
\nu_{\rm F} U T^2 / \epsilon_{\rm F}$.  However, away from the
transition point, the domain of applicability of the theory presented
earlier is determined by the following condition: $\Delta \ll \nu_{\rm
  F} U T^2/\epsilon_{\rm F}$.  At very low temperatures, the
quasiparticle distribution function cannot follow the changes of the
pairing field; to describe such a non-equilibrium regime, it is
necessary to use the Keldysh technique \cite{LO} or study the
equations of motion for the superfluid phase (Josephson equations)
coupled to the kinetic equations for the one particle distribution
function.  Unfortunately, the corresponding derivation is quite
cumbersome (and will be published elsewhere).  Below, we present only
qualitative results discussing several non-equilibrium regimes
possible in the system of interest:

It is natural to assume that by sweeping the magnetic field across the
resonance, we elevate the chemical potential of the bosonic component
relative to the chemical potential of the BCS condensate~\cite{NeqSC}.
The corresponding chemical potential difference $\delta$ yields an
effect similar to the effect of voltage in superconducting junctions.
Therefore, if $\delta \ll 2 \Delta$ the system should exhibit
oscillations associated with an {\em internal AC Josephson effect}:
\begin{equation}
\label{AC}
\dot{N}_{\rm b} = J \sin{\left( \delta t +
  \phi_0\right)}.
\end{equation}
These oscillations correspond to coherent tunneling back and forth
between the two quantum states (condensates of molecular bosons and
Cooper pairs). Due to the AC Josephson effect, the number of molecular
bosons would oscillate in real time and this behavior should in
principle be observable in experiment \cite{oscexp} with the help of
the radio-frequency spectroscopy technique \cite{PT}.
\begin{figure}[htbp]
\centering
\includegraphics[width=3.3in]{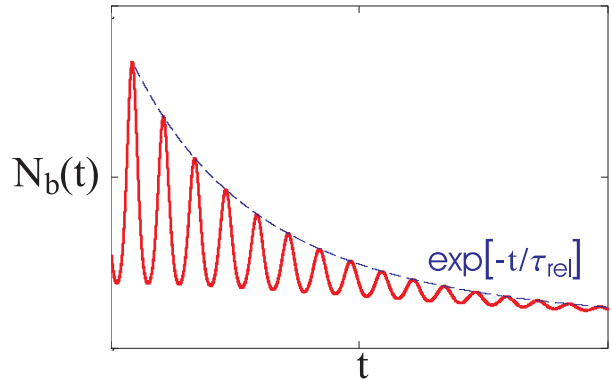}
\caption{ Real time dynamics of the number of molecular bosons
  in the Josephson regime: $\delta \ll 2\Delta$. The solid curve shows
  AC Josephson oscillations in the number of particles. The
  oscillations decay with time and the system tends to an equilibrium
  state, which generally depends upon the magnetic field sweeping
  rate.  } \vspace*{-0.2in}
\label{fig2}
\end{figure}

A different physical picture emerges when we consider the regime of a
highly excited molecular state with $\delta \gg 2 \Delta$. In this
limit, the molecular bosons quickly dissociate into one-particle
excitations and the Josephson effect disappears. In this case, one
would expect to observe Rabi oscillations, which were described in
details in Ref.~\cite{BLS} for a generic BCS problem.

The internal AC Josephson dynamics described above and the Rabi
oscillations of Ref.~\cite{BLS} are dissipationless, since
equilibration is possible only in the presence of relaxation processes
in the system of one-particle excitations. If the temperature is low
enough (keeping in mind that the word ``temperature'' has somewhat
ambiguous meaning in a non-equilibrium case) the relaxation to a final
equilibrium state is quite slow (see Fig.~2) and the corresponding
relaxation rate is likely to be exponentially small: $ \tau_{\rm
  rel}^{-1} \sim \tau_\varepsilon^{-1} \exp{\left[ - {{\rm min \,}
    \left\{ \delta,\Delta \right\} / T_{\rm i}} \right]}$, where
$T_{\rm i}$ is the initial temperature before the magnetic field
sweep.

Another important question concerns the final equilibrium state to
which the system relaxes at $t \to \infty$.  This state depends on the
final temperature $T_{\rm f}$ of the system. The latter is related to
the details of the magnetic field sweep.  By rapidly
(nonadiabatically) changing the magnetic field, one puts the system in
an excited state \cite{BLS,EI}; in the process of relaxation, the
system heats up being unable to drain the extra energy.  The
temperature change relates to the energy pumping rate $\dot{\delta}
\sim 2 \mu_{\rm B} \left({\partial B / \partial t}\right)$
\cite{BLS,EI}. If the final temperature is higher than the
superconducting transition temperature $T_{\rm f} > T_{\rm c}$, the
equilibrium phase of the system is a normal Fermi liquid state; in the
opposite limit $T_c < T_{\rm f}$, the system relaxes toward a
quasistationary state, which is either the Josephson coherent state
[if the system is large enough, $N > N_{\rm c}(T_{\rm f})$; see
Eq.~(\ref{pt}) and Fig.~1] or the Fock state [if $N < N_{\rm c}(T_{\rm
  f})$].

{\em Acknowledgements.} The author is very grateful to Anatoly Larkin
for his extremely helpful comments and to Donald Priour Jr. for
reading the manuscript. The author also acknowledges stimulating
discussions with Roman Barankov.  The work was supported by the
US-ONR, LPS, and DARPA.  \vspace*{-0.15in} \bibliography{fb}

\begin{thebibliography}{26}
\expandafter\ifx\csname natexlab\endcsname\relax\def\natexlab#1{#1}\fi
\expandafter\ifx\csname bibnamefont\endcsname\relax
  \def\bibnamefont#1{#1}\fi
\expandafter\ifx\csname bibfnamefont\endcsname\relax
  \def\bibfnamefont#1{#1}\fi
\expandafter\ifx\csname citenamefont\endcsname\relax
  \def\citenamefont#1{#1}\fi
\expandafter\ifx\csname url\endcsname\relax
  \def\url#1{\texttt{#1}}\fi
\expandafter\ifx\csname urlprefix\endcsname\relax\def\urlprefix{URL }\fi
\providecommand{\bibinfo}[2]{#2}
\providecommand{\eprint}[2][]{\url{#2}}

\bibitem[{\citenamefont{Levi}(2003)}]{PT}
\bibinfo{author}{\bibfnamefont{B.~G.} \bibnamefont{Levi}},
  \bibinfo{journal}{Phys. Today} \textbf{\bibinfo{volume}{56}},
  \bibinfo{pages}{20} (\bibinfo{year}{2003}).

\bibitem[{\citenamefont{Cho}(2003)}]{Cho}
\bibinfo{author}{\bibfnamefont{A.}~\bibnamefont{Cho}},
  \bibinfo{journal}{Science} \textbf{\bibinfo{volume}{301}},
  \bibinfo{pages}{750} (\bibinfo{year}{2003}).

\bibitem[{\citenamefont{Donley et~al.}(2002)\citenamefont{Donley, Claussen,
  Thompson, and Wieman}}]{oscexp}
\bibinfo{author}{\bibfnamefont{E.~A.} \bibnamefont{Donley}},
  \bibinfo{author}{\bibfnamefont{N.~R.} \bibnamefont{Claussen}},
  \bibinfo{author}{\bibfnamefont{S.~T.} \bibnamefont{Thompson}},
  \bibnamefont{and} \bibinfo{author}{\bibfnamefont{C.~E.}
  \bibnamefont{Wieman}}, \bibinfo{journal}{Nature}
  \textbf{\bibinfo{volume}{417}}, \bibinfo{pages}{529} (\bibinfo{year}{2002}).

\bibitem[{\citenamefont{Regal et~al.}(2004{\natexlab{a}})\citenamefont{Regal,
  Greiner, and Jin}}]{JinFC}
\bibinfo{author}{\bibfnamefont{C.~A.} \bibnamefont{Regal}},
  \bibinfo{author}{\bibfnamefont{M.}~\bibnamefont{Greiner}}, \bibnamefont{and}
  \bibinfo{author}{\bibfnamefont{D.~S.} \bibnamefont{Jin}},
  \bibinfo{journal}{\prl} \textbf{\bibinfo{volume}{92}},
  \bibinfo{pages}{040403} (\bibinfo{year}{2004}{\natexlab{a}}).

\bibitem[{\citenamefont{Zwierlein et~al.}(2004)\citenamefont{Zwierlein, Stan,
  Schunck, Raupach, Kerman, and Ketterle}}]{KetterleFC}
\bibinfo{author}{\bibfnamefont{M.~W.} \bibnamefont{Zwierlein}},
  \bibinfo{author}{\bibfnamefont{C.~A.} \bibnamefont{Stan}},
  \bibinfo{author}{\bibfnamefont{C.~H.} \bibnamefont{Schunck}},
  \bibinfo{author}{\bibfnamefont{S.~M.~F.} \bibnamefont{Raupach}},
  \bibinfo{author}{\bibfnamefont{A.~J.} \bibnamefont{Kerman}},
  \bibnamefont{and} \bibinfo{author}{\bibfnamefont{W.}~\bibnamefont{Ketterle}},
  \bibinfo{journal}{\prl} \textbf{\bibinfo{volume}{92}},
  \bibinfo{pages}{120403} (\bibinfo{year}{2004}).

\bibitem[{\citenamefont{Regal et~al.}()\citenamefont{Regal, Ticknor, Bohn, and
  Jin}}]{JinNature}
\bibinfo{author}{\bibfnamefont{C.}~\bibnamefont{Regal}},
  \bibinfo{author}{\bibfnamefont{C.}~\bibnamefont{Ticknor}},
  \bibinfo{author}{\bibfnamefont{J.~L.} \bibnamefont{Bohn}}, \bibnamefont{and}
  \bibinfo{author}{\bibfnamefont{D.~S.} \bibnamefont{Jin}},
  \bibinfo{howpublished}{Nature \textbf{424}, 47 (2003); M. Greiner, C. A.
  Regal, and D. S. Jin, Nature \textbf{426}, 537 (2003).}

\bibitem[{\citenamefont{Regal et~al.}(2004{\natexlab{b}})\citenamefont{Regal,
  Greiner, and Jin}}]{JinLifetime}
\bibinfo{author}{\bibfnamefont{C.~A.} \bibnamefont{Regal}},
  \bibinfo{author}{\bibfnamefont{M.}~\bibnamefont{Greiner}}, \bibnamefont{and}
  \bibinfo{author}{\bibfnamefont{D.~S.} \bibnamefont{Jin}},
  \bibinfo{journal}{\prl} \textbf{\bibinfo{volume}{92}},
  \bibinfo{pages}{083201} (\bibinfo{year}{2004}{\natexlab{b}}).

\bibitem[{\citenamefont{Regal and Jin}(2003)}]{Jina(B)}
\bibinfo{author}{\bibfnamefont{C.~A.} \bibnamefont{Regal}} \bibnamefont{and}
  \bibinfo{author}{\bibfnamefont{D.~S.} \bibnamefont{Jin}},
  \bibinfo{journal}{\prl} \textbf{\bibinfo{volume}{90}},
  \bibinfo{pages}{230404} (\bibinfo{year}{2003}).

\bibitem[{\citenamefont{Timmermans et~al.}(1999)\citenamefont{Timmermans,
  Tommasini, Hussein, and Kerman}}]{Feshbach}
\bibinfo{author}{\bibfnamefont{E.}~\bibnamefont{Timmermans}},
  \bibinfo{author}{\bibfnamefont{P.}~\bibnamefont{Tommasini}},
  \bibinfo{author}{\bibfnamefont{M.}~\bibnamefont{Hussein}}, \bibnamefont{and}
  \bibinfo{author}{\bibfnamefont{A.}~\bibnamefont{Kerman}},
  \bibinfo{journal}{Phys. Rep.} \textbf{\bibinfo{volume}{315}},
  \bibinfo{pages}{199} (\bibinfo{year}{1999}).

\bibitem[{\citenamefont{Nozi{\`{e}}res and Schmitt-Rink}(1985)}]{Noz}
\bibinfo{author}{\bibfnamefont{P.}~\bibnamefont{Nozi{\`{e}}res}}
  \bibnamefont{and}
  \bibinfo{author}{\bibfnamefont{S.}~\bibnamefont{Schmitt-Rink}},
  \bibinfo{journal}{J. Low. Temp. Phys.} \textbf{\bibinfo{volume}{59}},
  \bibinfo{pages}{195} (\bibinfo{year}{1985}).

\bibitem[{\citenamefont{Ohashi and Griffin}()}]{Griffin}
\bibinfo{author}{\bibfnamefont{Y.}~\bibnamefont{Ohashi}} \bibnamefont{and}
  \bibinfo{author}{\bibfnamefont{A.}~\bibnamefont{Griffin}},
  \bibinfo{howpublished}{Phys. Rev. Lett. \textbf{89}, 130402 (2002); Phys.
  Rev. A \textbf{67}, 1050 (2003)}.

\bibitem[{\citenamefont{Falco and Stoof}(2004)}]{Falco}
\bibinfo{author}{\bibfnamefont{G.~M.} \bibnamefont{Falco}} \bibnamefont{and}
  \bibinfo{author}{\bibfnamefont{H.~T.~C.} \bibnamefont{Stoof}},
  \bibinfo{journal}{\prl} \textbf{\bibinfo{volume}{92}},
  \bibinfo{pages}{130401} (\bibinfo{year}{2004}).

\bibitem[{\citenamefont{Stajic et~al.}()\citenamefont{Stajic, Milstein, Chen,
  Chiofalo, Holland, and Levi}}]{Stajic}
\bibinfo{author}{\bibfnamefont{J.}~\bibnamefont{Stajic}},
  \bibinfo{author}{\bibfnamefont{J.~N.} \bibnamefont{Milstein}},
  \bibinfo{author}{\bibfnamefont{Q.}~\bibnamefont{Chen}},
  \bibinfo{author}{\bibfnamefont{M.~L.} \bibnamefont{Chiofalo}},
  \bibinfo{author}{\bibfnamefont{M.~J.} \bibnamefont{Holland}},
  \bibnamefont{and} \bibinfo{author}{\bibfnamefont{K.}~\bibnamefont{Levi}},
  \bibinfo{howpublished}{cond-mat/0309329.}

\bibitem[{\citenamefont{Bohn}(2000)}]{Bohn}
\bibinfo{author}{\bibfnamefont{J.~L.} \bibnamefont{Bohn}},
  \bibinfo{journal}{\pra} \textbf{\bibinfo{volume}{61}},
  \bibinfo{pages}{053409} (\bibinfo{year}{2000}).

\bibitem[{\citenamefont{Holland et~al.}(2001)\citenamefont{Holland, Kokkelmans,
  Chiofalo, and Walser}}]{Holland}
\bibinfo{author}{\bibfnamefont{M.}~\bibnamefont{Holland}},
  \bibinfo{author}{\bibfnamefont{S.~J. J. M.~F.} \bibnamefont{Kokkelmans}},
  \bibinfo{author}{\bibfnamefont{M.~L.} \bibnamefont{Chiofalo}},
  \bibnamefont{and} \bibinfo{author}{\bibfnamefont{R.}~\bibnamefont{Walser}},
  \bibinfo{journal}{\prl} \textbf{\bibinfo{volume}{87}},
  \bibinfo{pages}{120406} (\bibinfo{year}{2001}).

\bibitem[{\citenamefont{Duine and Stoof}()}]{Stoof}
\bibinfo{author}{\bibfnamefont{R.~A.} \bibnamefont{Duine}} \bibnamefont{and}
  \bibinfo{author}{\bibfnamefont{H.~T.~C.} \bibnamefont{Stoof}},
  \bibinfo{howpublished}{cond-mat/0312254.}

\bibitem[{\citenamefont{Leggett}(2001)}]{Leggett}
\bibinfo{author}{\bibfnamefont{A.~J.} \bibnamefont{Leggett}},
  \bibinfo{journal}{\rmp} \textbf{\bibinfo{volume}{73}}, \bibinfo{pages}{307}
  (\bibinfo{year}{2001}).

\bibitem[{\citenamefont{Ambegaokar et~al.}(1982)\citenamefont{Ambegaokar,
  Eckern, and Sch$\ddot{\mbox{o}}$n}}]{AES}
\bibinfo{author}{\bibfnamefont{V.}~\bibnamefont{Ambegaokar}},
  \bibinfo{author}{\bibfnamefont{U.}~\bibnamefont{Eckern}}, \bibnamefont{and}
  \bibinfo{author}{\bibfnamefont{G.}~\bibnamefont{Sch$\ddot{\mbox{o}}$n}},
  \bibinfo{journal}{\prl} \textbf{\bibinfo{volume}{48}}, \bibinfo{pages}{1745}
  (\bibinfo{year}{1982}).

\bibitem[{\citenamefont{Abrikosov et~al.}(1990)\citenamefont{Abrikosov,
  Gor'kov, and Dzyaloshinski\^{\i}}}]{AGD}
\bibinfo{author}{\bibfnamefont{A.~A.} \bibnamefont{Abrikosov}},
  \bibinfo{author}{\bibfnamefont{L.~P.} \bibnamefont{Gor'kov}},
  \bibnamefont{and} \bibinfo{author}{\bibfnamefont{I.~Y.}
  \bibnamefont{Dzyaloshinski\^{\i}}}, \emph{\bibinfo{title}{Quantum field
  theorectical methods in statistical physics}} (\bibinfo{publisher}{Pergamon
  Press}, \bibinfo{address}{New York}, \bibinfo{year}{1990}).

\bibitem[{\citenamefont{Josephson}(1962)}]{Josephson}
\bibinfo{author}{\bibfnamefont{B.~D.} \bibnamefont{Josephson}},
  \bibinfo{journal}{Phys. Lett.} \textbf{\bibinfo{volume}{1}},
  \bibinfo{pages}{251} (\bibinfo{year}{1962}).

\bibitem[{\citenamefont{Chiofalo et~al.}(2002)\citenamefont{Chiofalo,
  Kokkelmans, Milstein, and Holland}}]{eqphi}
\bibinfo{author}{\bibfnamefont{M.~L.} \bibnamefont{Chiofalo}},
  \bibinfo{author}{\bibfnamefont{S.~J. J. M.~F.} \bibnamefont{Kokkelmans}},
  \bibinfo{author}{\bibfnamefont{J.~N.} \bibnamefont{Milstein}},
  \bibnamefont{and} \bibinfo{author}{\bibfnamefont{M.~J.}
  \bibnamefont{Holland}}, \bibinfo{journal}{\prl}
  \textbf{\bibinfo{volume}{88}}, \bibinfo{pages}{090402}
  (\bibinfo{year}{2002}).

\bibitem[{\citenamefont{Raghavan et~al.}(1999)\citenamefont{Raghavan, Smerzi,
  Fantoni, and Shenoy}}]{Raghavan}
\bibinfo{author}{\bibfnamefont{S.}~\bibnamefont{Raghavan}},
  \bibinfo{author}{\bibfnamefont{A.}~\bibnamefont{Smerzi}},
  \bibinfo{author}{\bibfnamefont{S.}~\bibnamefont{Fantoni}}, \bibnamefont{and}
  \bibinfo{author}{\bibfnamefont{S.~R.} \bibnamefont{Shenoy}},
  \bibinfo{journal}{\pra} \textbf{\bibinfo{volume}{59}}, \bibinfo{pages}{620}
  (\bibinfo{year}{1999}).

\bibitem[{\citenamefont{Larkin and Ovchinnikov}(1984)}]{LO}
\bibinfo{author}{\bibfnamefont{A.~I.} \bibnamefont{Larkin}} \bibnamefont{and}
  \bibinfo{author}{\bibfnamefont{Y.~N.} \bibnamefont{Ovchinnikov}},
  \bibinfo{journal}{JETP Lett.} \textbf{\bibinfo{volume}{39}},
  \bibinfo{pages}{681} (\bibinfo{year}{1984}).

\bibitem[{\citenamefont{{Eds., D. N. Langenberg} and Larkin}(1986)}]{NeqSC}
\bibinfo{author}{\bibnamefont{{Eds., D. N. Langenberg}}} \bibnamefont{and}
  \bibinfo{author}{\bibfnamefont{A.~I.} \bibnamefont{Larkin}},
  \emph{\bibinfo{title}{Nonequilibrium superconductivity}}
  (\bibinfo{publisher}{North-Holland}, \bibinfo{address}{Amsterdam},
  \bibinfo{year}{1986}).

\bibitem[{\citenamefont{Barankov et~al.}()\citenamefont{Barankov, Levitov, and
  Spivak}}]{BLS}
\bibinfo{author}{\bibfnamefont{R.~A.} \bibnamefont{Barankov}},
  \bibinfo{author}{\bibfnamefont{L.~S.} \bibnamefont{Levitov}},
  \bibnamefont{and} \bibinfo{author}{\bibfnamefont{B.~Z.}
  \bibnamefont{Spivak}}, \bibinfo{howpublished}{cond-mat/0312053.}

\bibitem[{\citenamefont{Eliashberg}()}]{EI}
\bibinfo{author}{\bibfnamefont{G.~M.} \bibnamefont{Eliashberg}},
  \bibinfo{howpublished}{JETP Lett. \textbf{11}, 114 (1970); G.~M.~Eliashberg
  and B.~I.~Ivlev in {\em Nonequilibrium Superconductivity},
  {Eds.,~D.~N.~Langenberg} and A.~I.~Larkin, North-Holland, Amsterdam, 1986,
  p.~211.}

\end{thebibliography}

\end{document}